# Imaging outside the box: Resolution enhancement in X-ray coherent diffraction imaging by extrapolation of diffraction patterns

Tatiana Latychevskaia,[1*] Yuriy Chushkin,[2] Federico Zontone[2] and Hans-Werner Fink[1]

[1]Physics Institute, University of Zurich, Zurich, Winterthurerstrasse 190, CH-8057, Switzerland

[2]The European Synchrotron Radiation Facility, 71 Avenue des Martyrs, 38000 Grenoble, France

[*]tatiana@physik.uzh.ch

**Abstract**

Coherent diffraction imaging is a high-resolution imaging technique whose potential can be greatly enhanced by applying the extrapolation method presented here. We demonstrate enhancement in resolution of a non-periodical object reconstructed from an experimental X-ray diffraction record which contains about 10% missing information, including the pixels in the center of the diffraction pattern. A diffraction pattern is extrapolated beyond the detector area and as a result, the object is reconstructed at an enhanced resolution and better agreement with experimental amplitudes is achieved. The optimal parameters for the iterative routine and the limits of the extrapolation procedure are discussed.

**Main text**

Conventionally, the resolution of an optical system is estimated by the Abbe criterion $R = \lambda/2\text{NA}$, where NA is the numerical aperture and $\lambda$ is the wavelength. According to this criterion, in lensless imaging, the sole limit of resolution (besides the wavelength) is the size of the detector. However, when dealing with coherent waves, the recorded far-field interference pattern can contain sufficient information to extrapolate the scattered waves beyond the detector area and thus to effectively increase *a posteriori* the resolution of reconstructed objects[1-2]. Such extrapolation has already been successfully demonstrated on light optical holograms and diffraction patterns[1-2], terahertz in-line holograms[3-4] and simulated diffraction patterns of crystalline samples[5]. The reported enhancement in resolution is at least twice the resolution obtained from non-extrapolated images[2,4].

A particular interest for extrapolation exists in coherent diffractive imaging[6], where the resolution is often limited by the size of the detector. In 1964, Harris speculated that the resolution of an optical system is not limited by the numerical aperture of the system but only by the experimental noise, because even a fraction of the detected spectrum is sufficient to uniquely restore the object details[7]. In 1974, Gerchberg addressed the problem of continuing a given segment of the spectrum of a finite object[8]. He restated the problem in terms of reducing a defined "error energy" and proposed an iterative computational procedure. As an example, Gerchberg considered an object consisting of two points, whose spectrum is a real-valued function. One year later, in 1975, Papoulis described the same algorithm but he inverted the domains[9]. However, the so-called Gerchberg–Papoulis algorithm cannot be directly applied to extrapolate diffraction patterns as it requires exact knowledge of a segment of a spectrum, including its phase distribution. Unfortunately, in a typical diffraction experiment only intensities are measured but the phase distribution is lost. When the Gerchberg–Papoulis algorithm is directly applied to an experimental diffraction pattern[9], it fails. In early report[10] a super-resolution algorithm was proposed in CDI based on zero-padding which is different from the extrapolation technique. As a solution to the problem of abrupt edges of a diffraction pattern, a slight extrapolation at the edges region was proposed where extrapolation however was limited to the edge regions and controlled by the weighting function; only a slight increase in resolution was achieved in this way[11]. Only recently successful extrapolations of diffraction patterns have been reported[2,5].

In this work we study how extrapolation can be applied to X-ray experimental data with missing information. We perform several reconstructions varying different parameters in the extrapolation routine. Some of the parameter values may seem arbitrary, but we have found that the resulting images are not affected significantly by their values. Thus, we present a study of the critical parameters that control the outcome of the extrapolation and reconstruction.

To study the behavior of the extrapolation algorithm we applied it to experimental X-ray diffraction data. The test sample is the logo of the ESRF patterned in a 220 nm thick tungsten film, which includes features of various sizes that are optimal for testing resolution. The scanning electron microscope (SEM) image of the sample and its diffraction pattern are shown in Fig. 1. The diffraction pattern was recorded at the ID10 beamline at the European Synchrotron

Radiation Facility (ESRF). A 7 keV (wavelength = 0.177 nm) coherent X-ray beam of $10 \times 10$ $\mu m^2$ in size with a flux of $8 \cdot 10^9$ photon/s illuminated the sample. The 2D diffraction data and the background with 10 s exposure time were taken using the Maxipix 2D detector having $516 \times 516$ pixels of 55 μm placed 5.16 m from the sample. The background image was subtracted from the measured 2D diffraction data to get the clean diffraction pattern. The pixel size in the object domain corresponds to 32 nm.

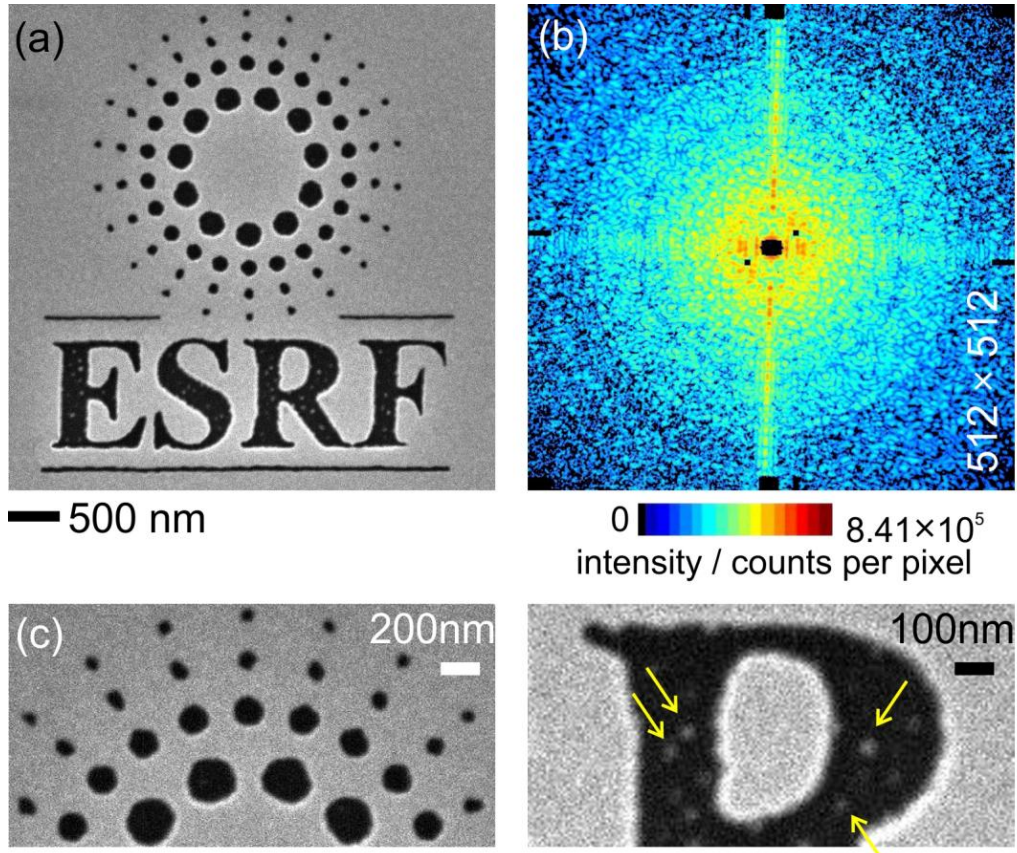

Fig. 1. The sample and its X-ray diffraction pattern. (a) Scanning electron microscope (SEM) image of the sample. (b) Its X-ray diffraction pattern shown in logarithmic intensity scale. The missing information is shown in black. (c) Selected regions of the SEM image. The yellow arrows indicate the features of the sample.

Information about the missing phase can be retrieved from the measured intensities sampled finer than the Nyquist frequency. The original diffraction pattern exhibits a size of $512 \times 512$ pixels, where 21,687 out of $512^2 = 262,144$ pixels, that are about 10%, are pixels with missing information. The conventional, non-extrapolation phase retrieval can be done using hybrid input-output (HIO) and error-reduction (ER) algorithms[12] and their modifications such as shrinkwrap algorithm[13], with the additional constraint applied to the object domain that the transmission function of the object must not exceed a certain threshold value[14]. It is worth noting that the exact details of obtaining a conventional reconstruction are not critical for the further extrapolation procedure, which only requires a stable reconstruction of the object, no matter how this reconstruction was achieved.

A total of 1000 reconstructions were obtained by applying the shrinkwrap algorithm for 2000 iterations. The quality of the retrieved complex-valued amplitudes in the detector plane was evaluated by calculating the mismatch between the measured and the iterated amplitudes, or the error:

$$E = \frac{\sum_{i,j=1}^{N} \left| G_{\exp}(i,j) - |G_{\mathrm{it}}(i,j)| \right|}{\sum_{i,j=1}^{N} |G_{\exp}(i,j)|}, \qquad (1)$$

where $G_{\exp}(i,j)$ are the experimentally measured amplitudes at the detector, $|G_{\mathrm{it}}(i,j)|$ are the iterated amplitudes, $i$ and $j$ are the pixel numbers $i,j=1...N$, and the missing pixels were excluded from the summation. 50 iterated distributions $G_{\mathrm{it}}(i,j)$ with the least error $E$ were selected and each one was stabilized with 1000 iterations of the error-reduction (ER) algorithm[12]. The details of the algorithms are provided in the supplemental material[15]. These 50 reconstructions were averaged; the result is shown in Fig. 2(a). The amplitude distribution in the detector plane with the recovered missing pixels is shown in Fig. 2(b). Fig. 2(c) shows the magnified regions of the object reconstruction to compare them to the same regions in the SEM image shown in Fig. 1(c).

The quantitative estimation of the resolution is done by calculating the Phase Retrieval Transfer Function (PRTF)[16-17]:

$$\mathrm{PRTF}(u) = \frac{|\langle G_{\mathrm{it}}(u) \rangle|}{|G_{\exp}(u)|}, \qquad (2)$$

where $u = \sqrt{(i-N/2)^2 + (j-N/2)^2}$ is the spatial frequency coordinate, shown in Fig. 2(b), and $\langle...\rangle$ denotes averaging over the complex-valued iterated amplitudes; the missing pixels are excluded from averaging. The PRTF estimates the relation of the recovered amplitudes to the experimentally measured amplitudes. Figure 2(d) exhibits PRTF for an average of 50 reconstructions.

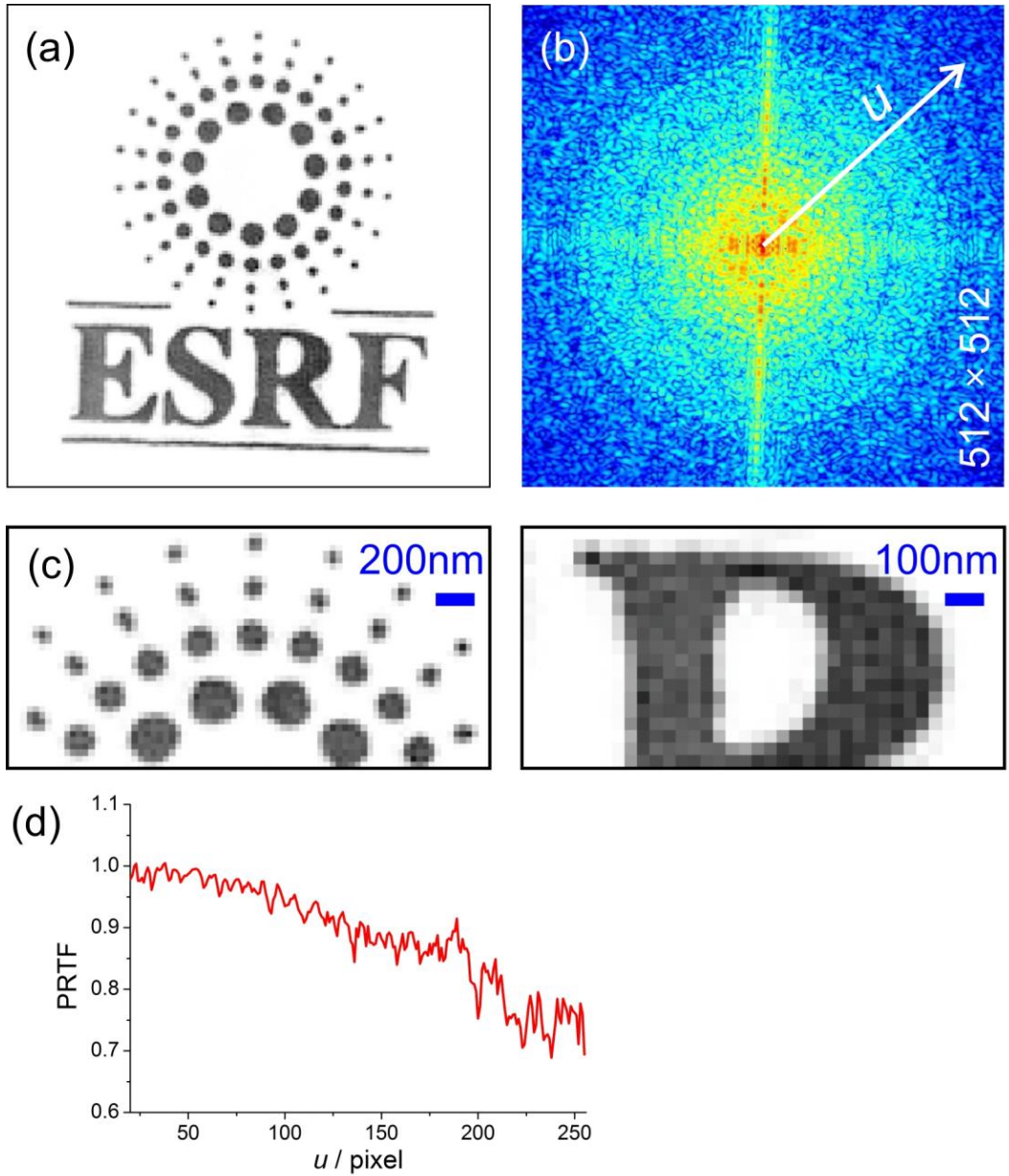

Fig. 2. Reconstruction of the sample obtained by conventional phase retrieval. (a) The central 160 × 160 pixels part of the reconstruction. (b) 512 × 512 pixels

diffraction pattern with retrieved dead pixels; shown in logarithmic intensity scale. (c) Selected regions of the reconstructed object distribution. (d) Calculated PRTF as a function of spatial frequency $u$ in pixels; the coordinate $u$ is indicated in (b).

To obtain the image of the test object with higher resolution the diffraction pattern must be extrapolated. Each of the 50 retrieved complex-valued distributions is extrapolated individually. To extrapolate the signal beyond the detector area we applied the iterative procedure as described elsewhere[1-2]. During the extrapolation, the pixel size in the detector (Fourier) domain does not change, but the number of pixels is increased. Thus, the reconstructed object area size remains the same as obtained from a conventional reconstruction of the experimental diffraction pattern, but it is sampled with an increased number of pixels.

The extrapolation begins in the Fourier domain, where the complex-valued wavefront distribution obtained by the conventional phase retrieval routine is padded with random complex-valued numbers up to $1024 \times 1024$ pixels. The amplitude of the padding signal is selected to be randomly distributed ranging from 0 to $A$, where $A$ is approximately equal to the amplitude level at the rim of the measured diffraction pattern, $A = 100$. The phase of the padding signal is randomly distributed from $-\pi$ to $+\pi$.

The iterative routine is based on the ER algorithm[12]. During iterations, the object distribution is multiplied with the object support which sets the pixel values outside the masked region to zero. The object support $\gamma$ is obtained by sampling of the reconstructed object area with $1024 \times 1024$ pixels, followed by a convolution it with a two-dimensional Gaussian distribution with $\sigma = 3$ and a threshold applied to the amplitude of the result at 15% of its maximum. The object support is updated after each iteration.

The object is assumed to be real-valued and thus the phase distribution of the object was set to 0. The amplitude distribution was set to not exceed a certain threshold. At pixels where the amplitude exceeds the threshold it is set to the threshold value. In the Fourier domain, the amplitudes are replaced with the experimental values but updated at the missing pixels and in the padding region; the phase distribution was updated after each iteration. A total of 500 iterations were applied. The final reconstruction was obtained by alignment and averaging of reconstructions of those 50 extrapolated diffraction patterns. The square amplitude of the Fourier transform of the final reconstruction gave provided the extrapolated diffraction pattern.

In the following we investigate the effect of the threshold in the object domain, the amplitude of initial padding and the size of the extrapolation area on the reconstructed images.

We studied the effect of the threshold in the object domain; the results are shown in Fig. 3. During the conventional phase retrieval the amplitude in the object domain did not exceed 2. According to the Parseval theorem, the total intensity distribution in the object domain must remain unchanged; this causes the amplitude of 2 in the case of $512 \times 512$ pixels to be reduced to the amplitude of 0.5 in the case of $1024 \times 1024$ pixels. Therefore, we selected the three different thresholds: 0.3, 0.35 and 0.4, the extrapolated diffraction patterns at these thresholds are presented in Fig. 3(a). Magnified sections of the extrapolated diffraction patterns are shown in Fig. 3(b). The extrapolated part appears to be a smooth continuation of the experimental diffraction pattern, and no border between the two parts can be distinguished, except for the threshold = 0.40.

The magnified parts of the reconstruction are shown in Fig. 3(c). In the reconstructed sample distribution, the various shapes of the circles can be well related to those in the SEM image. These reconstructions are obviously superior when compared to the reconstruction obtained by the conventional phase retrieval shown in Fig. 2(c). At a threshold = 0.35, there are indications of the spots of approximately 40 nm in diameter recovered inside the letters, which demonstrates the resolution enhancement, compare the features indicated in Fig. 1(c) with the features indicated in Fig. 3(c). However, at a higher threshold = 0.4, there are wavy artifacts showing up inside the letters. Thus, a correctly selected threshold is important for recovering the internal structure of the sample.

The calculated PRTFs are shown in Fig. 3(d). Unlike in Fig. 2(d), here all the PRTFs are higher than 0.80 and do not decrease as a function of $u$. This means that the recovered amplitudes match the measured amplitudes better. A similar observation has been reported in the case of extrapolating holograms[1], when the retrieved amplitudes within the hologram matched the experimental amplitudes better once the extrapolation had been applied. This can be explained by the fact that the constraint of a diffraction pattern being confined within a certain volume is unnatural. Removing that constraint thus improves the match between the experimentally measured and phase retrieved amplitudes. The PRTF at a threshold=0.35 exhibits values that are closer to 1 than the values of the PRTFs at a thresholds 0.30 and 0.40.

Thus, both the visual inspection of the three reconstructions and the calculated PRTFs indicate that the reconstruction obtained at the threshold=0.35 is the best match of the sample distribution. In the following analysis, this threshold = 0.35 in the object domain is selected.

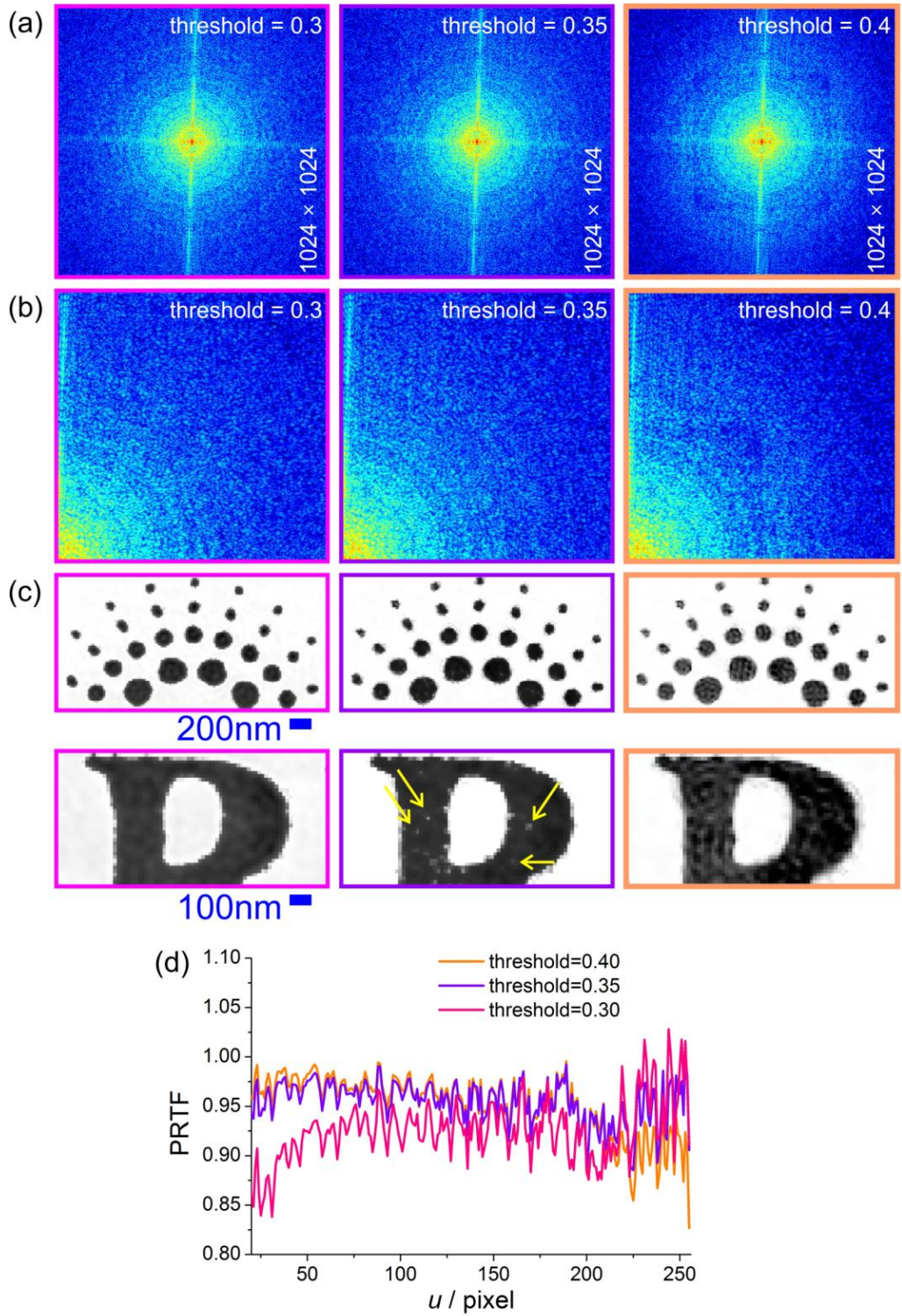

Fig. 3. Extrapolation of a diffraction pattern at different thresholds in the object domain. (a) Extrapolated to 1024 × 1024 pixels diffraction patterns at the thresholds in the object domain of 0.30, 0.35 and 0.40. (b) Magnified top right

regions of the extrapolated diffraction patterns. (c) Selected regions in the reconstructed object distribution related to the selected regions in the SEM image in Fig. 1. The yellow arrows indicate the features of the sample. (d) The calculated PRTF as a function of spatial frequency $u$ in pixels.

The reconstructions at different complex-valued random padding at the first iteration, $A = 0$ and $A = 1000$, are shown in Fig. 4(a). Both reconstructions appear visually to be of the same quality. The related PRTFs, shown in Fig. 4(d), are also almost identical. From this we can deduce that the amplitude of the random complex-valued padding at the first iteration is not very important for the final result, but a difference can be seen in the speed of convergence: properly selected $A$ ensures a fast arrival at the stable reconstruction.

The padding with zeros at the first iteration ($A = 0$) should not be confused with zero-padding. In zero-padding, the spectrum is padded with zeros, whereas in extrapolation, the initial zeros eventually turn themselves into non-zero values. Just zero-padding of a diffraction pattern or a hologram does not increase the numerical aperture of the optical system, and therefore no resolution enhancement is achieved[2]. However, zero-padding enhances sampling of the reconstructions[18-20] and represents "ideal" interpolation.

We also studied the extrapolation to 1536 × 1536 pixels and to 2048 × 2048 pixels areas. The amplitude of the initial random padding was selected to be $A = 100$. The thresholds in the object domain were selected to be 0.156 and 0.085 for extrapolation to 1536 × 1536 pixels and 2048 × 2048 pixels, respectively. Figure 4 (b) shows the results of extrapolations to 1536 × 1536 pixels and to 2048 × 2048 pixels. The related reconstructions shown in Fig. 4(c) show that the outer contours of the features and the distributions of spots inside the letters become more pronounced, see for example, the upper part of the letter "R". Calculated PRTFs plotted in Fig. 4(d) are very similar for 1536 × 1536 pixels and 2048 × 2048 pixels extrapolated diffraction patterns. Apparently, the efficiency of the retrieval of the amplitudes in the Fourier domain stagnates. This leads to the conclusion that extrapolation to an even larger area will be meaningless.

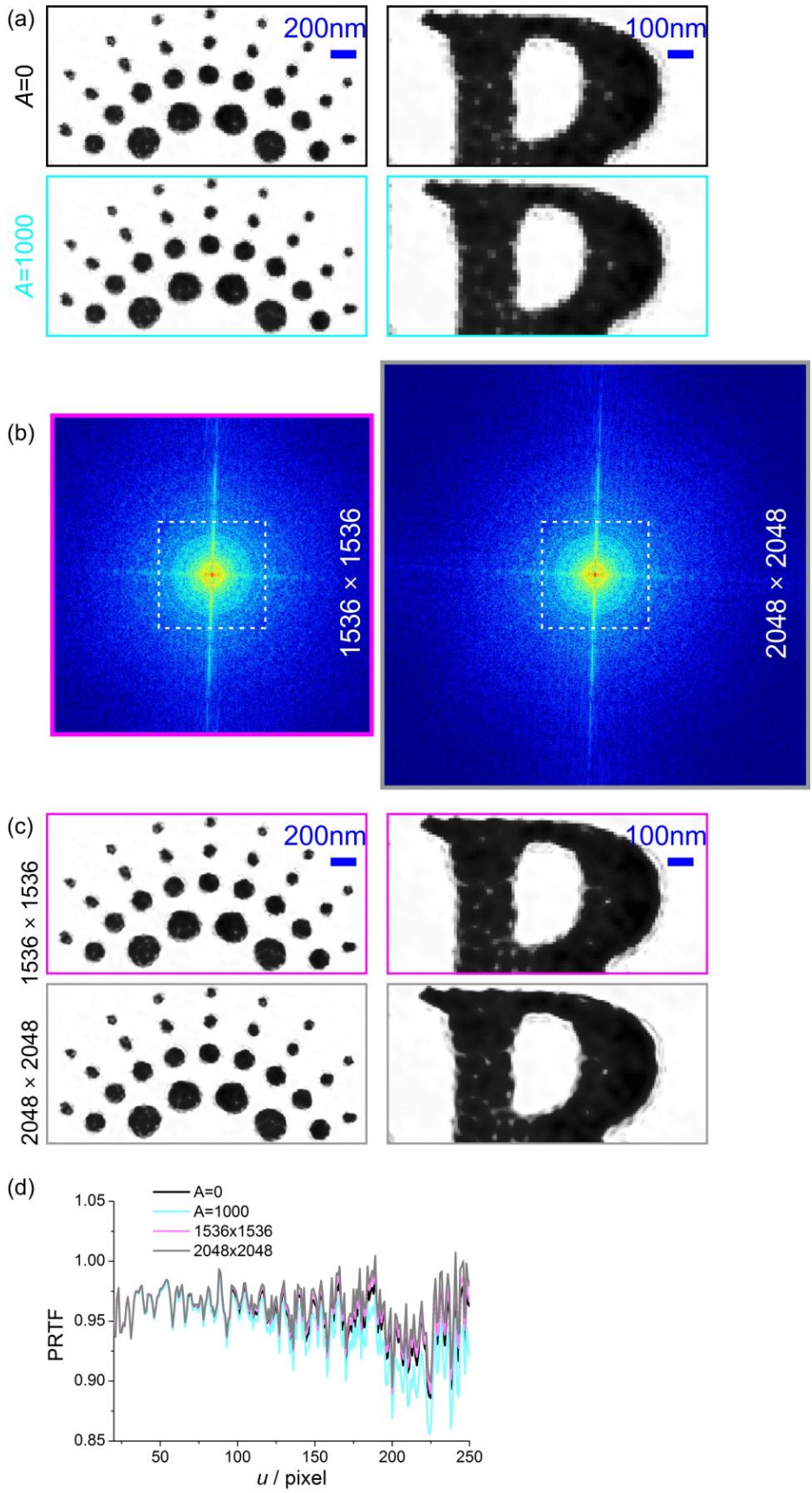

Fig. 4. Reconstructions of extrapolated diffraction pattern. (a) Selected regions of the reconstructed object distributions obtained with $A = 0$ and $A = 1000$. (b) Extrapolated diffraction patterns to $1536 \times 1536$ pixels and $2048 \times 2048$ pixels and (c) Selected regions of the object reconstructed from the related diffraction patterns. (d) Calculated PRTF as a function of spatial frequency $u$ in pixels.

**Conclusions**

We have applied an extrapolation procedure to a experimental X-ray diffraction pattern and studied the optimal conditions for the best reconstructions. We found that the amplitude of the random complex-valued padding distribution at the first iteration is not very important for the final result, but when properly selected (to match the amplitude at the rim of the experimental diffraction pattern), it leads to a faster convergence to the sample reconstruction. The important parameter turned out to be the threshold applied to the amplitude of the object distribution. It controls the quality of the reconstruction. When selected too low it leads to opaque reconstructions, and when selected too high it leads to wavy artifacts within the object distribution. We have selected the threshold to be constant but we believe that by further optimization of this constraints, there is a potential for further resolution enhancement by extrapolation.

The quantitative measure of the achieved phase retrieval, the PRTF, shows that in the extrapolated diffraction patterns the retrieved amplitudes match better the experimentally recorded amplitudes than the amplitudes obtained by conventional phase retrieval. This leads to the conclusion that the constraint that a diffraction pattern should be limited only to the detector area, therefore with waves beyond the detector abruptly turned into zeros, is unphysical. When this constraint is removed, waves are allowed to be non-zero *beyond* the detector area and these waves are then recovered more precisely. As a result, the amplitudes of these recovered waves match better the measured amplitudes *within* the detector area.

**Acknowledgments**

The University of Zurich is gratefully acknowledged for its financial support. We acknowledge the European Synchrotron Radiation Facility (ESRF) for its provision of the synchrotron radiation facilities. ESRF logo credit the European Synchrotron Radiation Facility. We kindly acknowledge Irina Snigireva from the ESRF for the supplementary SEM analysis.